\documentstyle[12pt,epsf]{article}
\begin{document}

\title{Fine "mist" vs large droplets in phase separated manganites}

\author{L.~Khomskii\\
St~Edmund's College, Cambridge University,\\Cambridge CB3~0BN, UK\\\\
D.~Khomskii\\
Laboratory of Solid State Physics,\\Groningen University,
Nijenborgh 4, 9747 AG Groningen,\\The Netherlands}
\date{}
\maketitle

\begin{abstract}
\indent The properties of phase-separated systems, e.g.\ manganites,
close to a I~order phase transition between charge-ordered insulator
and ferromagnetic metal, are usually described by the percolation
picture.  We argue that the correlated occupation of metallic sites
leads to the preferential formation of larger metallic clusters,
which explains the often observed inverse, or ``overshot'' hysteresis
in manganites (when the resistivity with increasing temperature is
larger than with decreasing~$T$).  It also explains the recently
discovered thermal cycling effect in manganites.  Thus in treating
this and similar systems in percolation picture, not only the total
concentration of metallic phase, but also the distribution of
metallic clusters by shape and size may significantly influence the
properties of such systems.
\end{abstract}

\medskip
\noindent\kern4emPACS numbers: 71.10.w, 75.30.Kz

\medskip\bigskip


Phase separation seems to be the generic feature of doped strongly
correlated systems such as manganites La$_{1-x}M_x$MnO$_3$ ($M=\rm Ca$,
Sr).  It is observed in many situations experimentally and is
obtained in many theoretical models~\cite{dagotto,nagaev,khomskii},
both at low doping range ($x<1$) and close to a half-doped case
($x\sim0.5$). Apparently one can speak of two different types of
phase separation: microscopic phase separation, which is most often
discussed by theoreticians and which is observed e.g.\ by the
small-angle neutron scattering~\cite{hennion}, and the large-scale,
macroscopic phase separation.  The later type is typically met close
to a first-order phase transition and leads to a percolation-like
behaviour of the system.

Two unusual effects were observed recently in studying the behaviour
of certain manganites close to a I~order insulator--metal transition
--- in (PrCa)MnO$_3$ \cite{lorenz},
Pr$_{0.5}$Ca$_{0.5}$(MnCr)O$_3$ \cite{mahendiran} and in some
others~\cite{babushkina}. In these systems there apparently occurs
with decreasing temperature a transition from a charge-ordered (CO)
insulator to a ferromagnetic metallic (FM) phase, accompanied by a
sharp drop of resistivity. This drop has a large hysteresis. But with
increasing temperature from the FM phase an inverse, or ``overshot''
hysteresis was observed in~\cite{lorenz,mahendiran,babushkina},
schematically shown in fig.~1. The nature of this behaviour was not
clarified; there were even suggestions~\cite{lorenz} that there exist
two different CO phases, one appearing with decreasing temperature,
and another --- when the temperature increases.

Another unusual phenomenon was found in one of these systems,\break
Pr$_{0.5}$Ca$_{0.5}$(MnCr)O$_3$ \cite{mahendiran}: when after the
first decrease of the temperature it was increased and then the cycle
was repeated, the resistivity, having the behaviour like the one shown
in fig.~1, in each following cycle became higher and higher (if
temperature was not increased beyond the shaded region of fig.~1).
In some cases after several cycles the resistivity became insulating
down to the lowest temperatures.  Simultaneous magnetic measurements
did not show any significant decrease of the total magnetization,
i.e.\ the total fraction of the FM phase did not decrease strongly with
such ``training.''

\begin{figure}
\centerline{\epsfbox{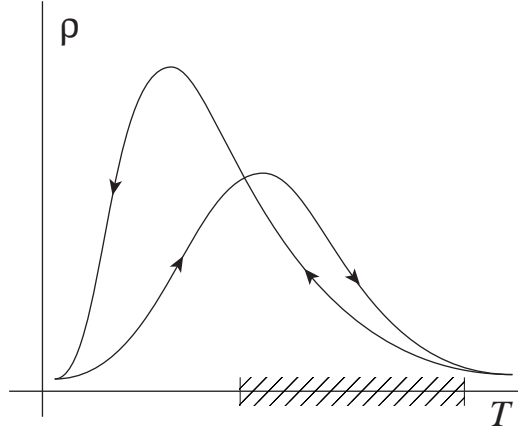}}
\caption{Schematic behaviour of resistivity in certain
manganites close to CO--FM transition [5--7].  Shaded is the region of
``overshot'' hysteresis.}
\end{figure}

We want to suggest here a simple explanation of these effects,
introducing the notion generalizing the standard percolation picture.
This idea was first put forth in 1999~\cite{khomskii2}; closely
related ideas were discussed recently in~\cite{burgy}.

The main idea is the following.  When there occurs a first order
phase transition to a metallic phase with decreasing temperature,
there appear FM droplets in a CO insulating matrix.  With
further decrease of temperature they grow and start to coalesce until
a percolation limit is reached, after which the system behaves as a
metallic one.  During this process the metallic droplets are first
formed close to some nucleation centres, so that there appear many
very small droplets --- like a fog on a cold evening.  With further
decrease of temperature these droplets grow, and, as is well known,
bigger droplets grow faster, gradually ``consuming'' smaller ones.
This is caused by the larger vapour pressure above droplets with
smaller curvature~\cite{landau}, or, in other words, by the tendency
to reduce total surface energy. Finally the FM phase occupies
(almost) the whole sample, which occurs at low temperatures.

With the following increase of the temperature the FM phase gradually
starts to ``evaporate'', but this process is accompanied by
hysteresis --- resistivity is initially lower than at the first
decrease of temperature.  However after a percolation metallic
cluster is broken, there may occur an inverse situation --- inverse,
or ``overshot''  hysteresis, at which the resistivity is higher than
at the first downward run.  This is a natural consequence of
different distribution of FM clusters by size and shape during
decreasing and increasing temperature: whereas in going from the CO
insulator phase we create a lot of small FM clusters (fine ``fog''),
in the opposite process, when we increase temperature starting from
the FM phase occupying large part of the sample (big FM ``pools'') ,
these big droplets would survive even at high temperature.  Thus at a
given temperature (in the ``shaded'' region of fig.~1) the total {\it
volume} of a FM phase may still be the same, but the {\it
distribution}, the typical sizes of these FM clusters would be
different: a lot of small droplets with
decreasing~$T$, and much smaller number of bigger droplets
when we increase temperature from the FM phase.  This would
naturally lead to an increase of resistivity in a reverse run --- an
overshot hysteresis: not only would FM droplets be bigger, but also
insulating regions between them would increase, which would lead to
larger resistivity.

\begin{figure}
\centerline{\epsffile{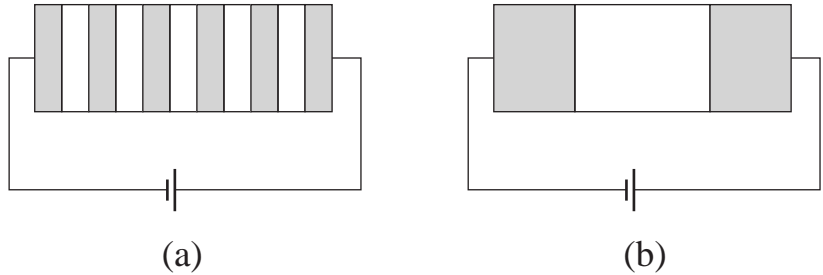}}
\caption{Schematic picture illustrating the dependence of resistivity
on the distribution of metallic regions (shaded) by size.}
\end{figure}

One can illustrate this conclusion on a simple
picture shown in fig.~2, in which we substituted random distribution
of FM and CO phases by a regular stripe-like structure.  A fine
``fog'' (large number of small FM and CO regions), realized with the
decreasing $T$, is modelled by the situation of fig.~2$a$, and a
situation which should be realized with the temperature increase is
illustrated in fig.~2$b$.  One immediately sees that the resistivity
in the first case is given by the expression
\begin{equation}
\rho_{1a}=\rho_0\,n\,e^{V/kT}\label{eq1a}
\end{equation}
where $n$ is the number of insulating barriers (white stripes in
fig.~2$a$), and $V$ is the value of each of these barriers (which for
simplicity we take equal).

On the other hand, in the case of fig.~2$b$, instead
of having $n$ small barriers, we have smaller number --- in a
limiting case only one barrier, but with the width
$n$ times bigger.  As a result we would get the
resistivity
\begin{equation}
\rho_{1b}=\rho_0\,e^{nV/kT}\label{eq1b}
\end{equation}
--- much larger than that given by Eq.~(\ref{eq1a}).

Of course this model is strongly oversimplified, and in reality the
difference between resistivities would be much smaller due to random
distribution of different regions by size, shape and position;
but the physics of ``overshot hysteresis'' may be explained by the
picture described above.  Thus when considering the percolation
conductivity, we have to take into account not only the relative
volume, occupied by one or another phase, but also the {\it
distribution} of these phases by the size.  This is usually not done
in a standard treatment of percolation; but, as we argued above, this
may be a very important factor~\cite{comment}.

We checked this picture by a computer simulation.  We modelled the
percolation in phase-separated manganites by first randomly putting
the ``metallic atoms'' (black points) on a $200\times200$ square
lattice.  The resulting distribution of metallic clusters for certain
concentration $n_0$, smaller than the percolation threshold
$n_c\sim0.59$~\cite{stauffer} (here for $n_0=0.125$) is shown in
fig.~$3a$. This distribution is on the average the same for increasing
and for decreasing~$n$.

\begin{figure}
\centerline{\epsffile{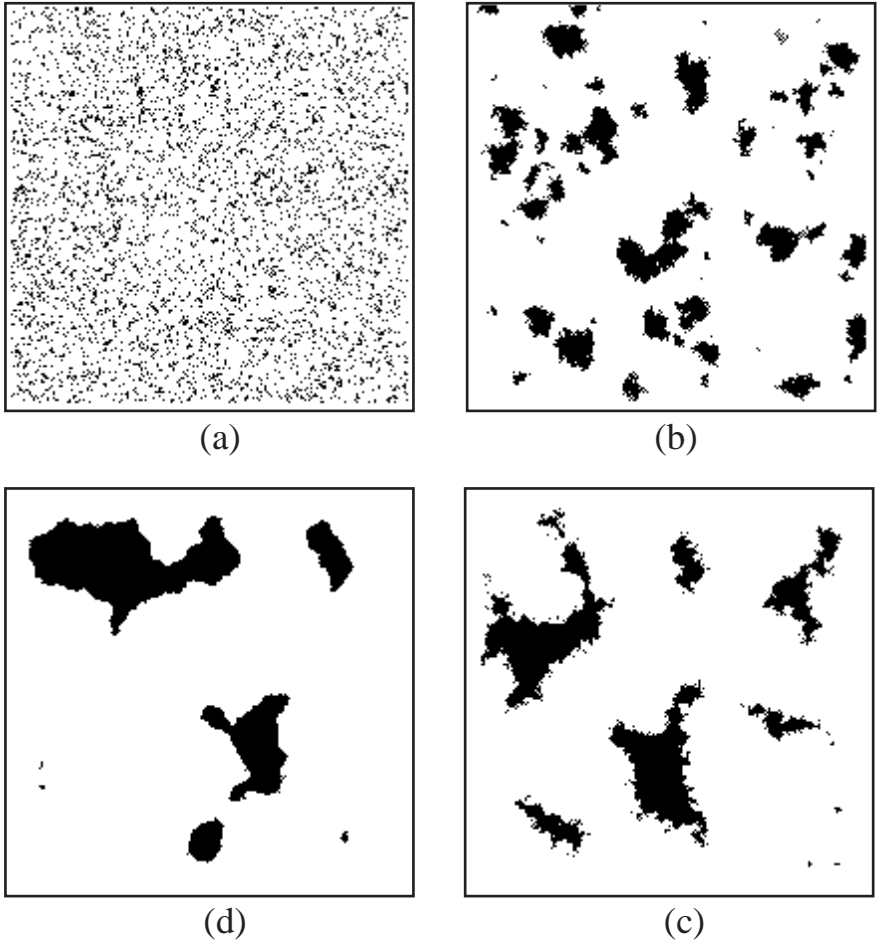}}
\nobreak\bigskip
\caption{Distribution of metallic clusters (black) for the filling
$n_0=0.125$ reached  at random occupation (3$a$) and for correlated
occupation of metallic sites (3$b$--3$c$), see text.  Fig.~3$b$
corresponds to a state reached by increasing $n$ from 0 to $n_0$, and
fig.~3$c$ --- by first increasing $n$ still further to $\sim0.75$
and then reducing it back to $n_0$.  Fig.~3$d$ shows typical
distribution of FM clusters at the same $n_0$ reached with
annealing.}
\end{figure}

To model the physical situation described above (the preferable
formation of large clusters) the algorithm was modified in such a way
that the probability of adding new metallic atom at a certain cite is
larger when there are already occupied sites adjacent to it (i.e.\
the probability to occupy the site is the larger, the more
neighbouring sites are occupied).  The resulting
structures are shown in fig.~3$b$, 3$c$.  Fig.~3$b$ shows the
distribution of occupied sites at $n_0=0.125$ reached by increasing
occupation $n$ from zero with correlated occupation as explained
above.  Fig.~3$c$ shows the distribution at the same concentration
$n_0$ as in fig.~$3b$, but reached by first increasing $n$ from
the situation of fig.~3$b$ to $n\sim0.75$ (above percolation
threshold) and then decreasing $n$ down to 0.125; in reducing $n$ we
used the same algorithm as when increasing it, i.e.\ the probability
to remove an atom from a given site is larger when there are fewer
occupied sites nearby.

As we see by comparing fig.\ 3$a$ with  3$b$ and 3$c$, the resulting
distribution of metallic sites at the same total concentration (here
0.125) depends on whether we have random or correlated
percolation: the clusters are bigger for correlated occupation. But
more interestingly, the resulting distribution also depends on
history: in accordance with our general expectations, for correlated
occupation we indeed obtain many small clusters with increasing
$n$ (or decreasing temperature), fig.~3$b$, and smaller number of
bigger clusters, with bigger insulating barriers between them, with
decreasing $n$ (increasing temperature), fig.~3$c$.

We can also add yet another ingredient in our computer modelling,
imitating annealing: after several steps of adding particles, we
allowed for their redistribution, removing and adding particles in
the same correlated fashion, but keeping their total number fixed.
This leads to some ``rounding off'' of the clusters, whose boundaries
become smoother, and this annealing somewhat enhances the tendency
described above: that for correlated occupation we obtain, on the
average, larger droplets, see fig.~3$d$, and they become even larger
at the reverse process of decreasing~$n$.

One also sees in these simulations that the percolation limit $n_c$
itself does not depend on whether we increase or decrease $n$, even
for correlated occupation; but the value of $n_c$ seems to decrease
somewhat, from $\sim0.593$ for the usual percolation~\cite{stauffer}
to about $0.576$ for correlated occupation.

Thus our computer modelling confirms our general physical arguments
and shows that in a more realistic picture of phase separation, which
takes into account correlation in occupation of particles and which
leads to the preferential formation of bigger clusters, the resulting
picture depends on the history: we have many small clusters (fine
``fog'') with increasing the fraction of metallic sites (decreasing
temperature), and smaller number of bigger clusters with
increasing~$T$.  As argued above, on the insulating side of the
transition  (below percolation threshold) this would lead to an {\it
increase} of resistivity (inverse, or ``overshot'' hysteresis) which
may explain the experimental observations
of~\cite{lorenz,mahendiran,babushkina}.  Thus, the correlated
occupation of sites makes the system ``more insulating'' on the
insulating side of the percolation transition and ``more metallic''
on the metallic side (the sharpening of the transition
on the metallic side was also seen in the calculations
of Ref.~\cite{burgy}). The most interesting feature of this picture
is the dependence of it on the thermal history, shown above.

The picture suggested above may also explain the ``training'' effect
observed in~\cite{mahendiran}.  Indeed, one may expect that after the
first cycle not all small droplets disappear.  But with further
cycling the larger and larger droplets will be formed, eventually with
larger insulating barriers between them, which can lead to the
behaviour observed in~\cite{mahendiran}.  The requirement is that the
upper temperature during cycling should fall within the ``shaded''
region of fig.~1$a$ (and should not exceed it) so that the large
droplets, which would serve as condensation centres during the next
cycle, should not disappear.

Our computer modelling also confirms these qualitative
considerations.  In fig.~4$a$--$d$ we show typical results of the
distribution of FM (black) regions obtained after cycling.
The
procedure was first to increase $n$ from 0 to certain $n_0<n_c$ (here
again $n_0=0.125$), and then cycling $n$ several times between $n_0$
and $n\sim0.75$.  All the time we used the same algorithm as before
(probability of adding and removing particles depending on the number
of occupied neighbours). We see that, indeed, with increasing number
of cycles the number of FM droplets for the same $n_0$ decreases, and
their size and distance between them increase, which, according to the
arguments leading to Eqs.\ (\ref{eq1a})--(\ref{eq1b}), gives an
increase of resistivity with ``training'', even for the same
concentration of the FM phase.  This can explain the experimental
results of Mahendiran et al.~\cite{mahendiran} (similar behaviour is
also seen in the data of~\cite{lorenz}).

\begin{figure}
\centerline{\epsffile{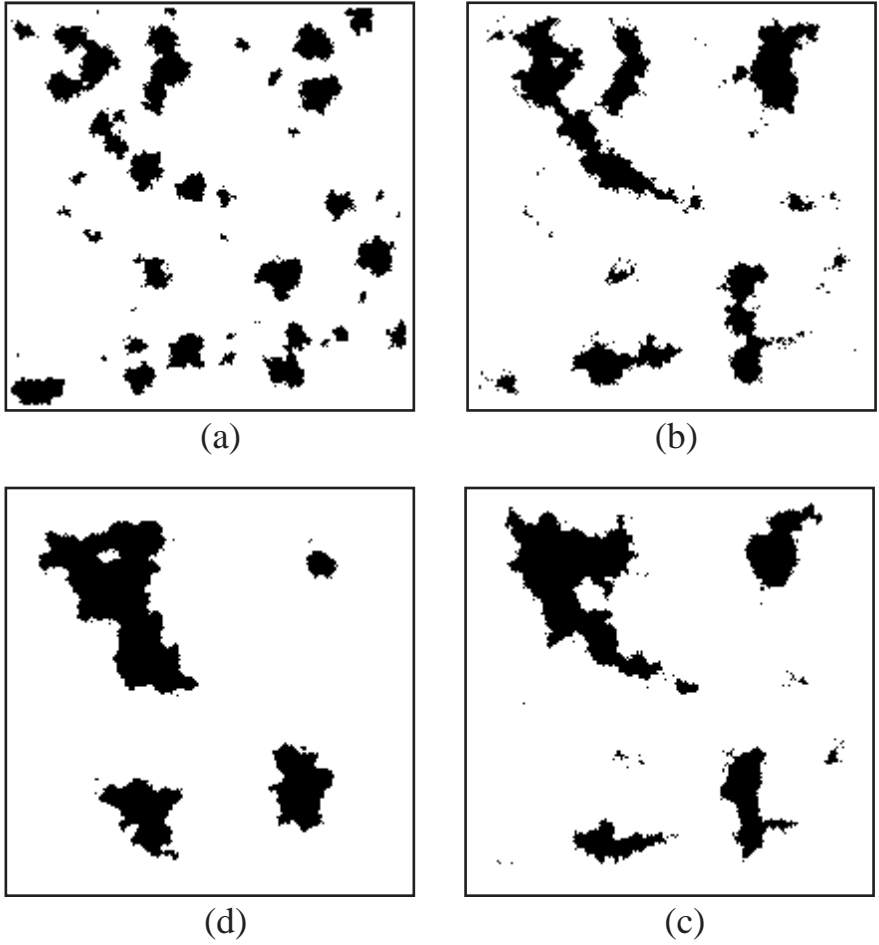}}
\nobreak\bigskip
\caption{The effect of thermal cycling.  The occupied metallic
clusters (black) for $n_0=0.125$ after first increasing $n$ from 0
to $n_0$ ($a$), further increasing n to $0.75$ and then decreasing it
back to $n_0$ ($b$) and after 2 ($c$) and 5 ($d$)
cycles. (Figs.~4$a$ and 4$b$ correspond to figs.~3$b$ and 3$c$.)}
\end{figure}

Summarizing, we proposed that the properties of inhomogeneous systems
like some manganites close to an insulator--metal transition may be
explained if we add to the conventional percolation picture another
ingredient --- that not only the net concentration of metallic
phase, but also the distribution of these metallic inclusions by size
and shape may be different, which may strongly influence the
properties. We argue that the metallic droplets formed with
decreasing temperature take the form of a fine ``fog'' --- a lot of
small droplets formed at many different nucleation centres, whereas
with increasing temperature, going from the metallic state,
predominantly large metallic droplets survive. The picture we propose
seems to be quite natural and agrees with what we know from other
fields of physics and even from our everyday experience; it is also
confirmed by our computer modelling. It can explain the ``inverse
hysteresis'' and the change of properties during thermal cycling,
observed in some manganites in the inhomogeneous phase close to an
insulator--metal transition.

The general conclusion is that when treating the properties of
inhomogeneous systems in percolation picture, not only the total
fraction of one or another phase, but also distribution of these
phases by size and shape may be crucial.  It would be very
interesting to verify the proposed picture experimentally,
e.g.\ by small angle neutron scattering or by light scattering in
manganites during thermal cycling.

We are grateful to E.~Dagotto and M.~Mostovoy for useful discussions
and to Th.~Lorenz and R.~Mahendiran for informing us of their
experimental results.  This work was supported by the Netherlands
Foundation for the Fundamental Study of Matter (FOM).


\begin{thebibliography}{99}
\bibitem{comment}The physics invoked above is actually well known
from our everyday experience: in the right weather conditions, when
the temperature drops, e.g.\ in the evening, an overcooled water
vapour can often form fog; but after big water droplets are finally
formed, they survive when the temperature increases (e.g.\ in the
morning), and the fog rarely forms, we rather have dew-drops instead.
\end{thebibliography}
\end{document}